\documentclass{article}
\usepackage[utf8]{inputenc}
\usepackage[margin=1in]{geometry}
\usepackage{graphicx} 
\usepackage{subcaption}  
\usepackage{upgreek}
\usepackage{graphicx}%
\usepackage{multirow}%
\usepackage{amsmath,amssymb,amsfonts}%
\usepackage{amsthm}%
\usepackage{mathrsfs}%
\usepackage[title]{appendix}%
\usepackage{xcolor}%
\usepackage{textcomp}%
\usepackage{manyfoot}%
\usepackage{booktabs}%
\usepackage{algorithm}%
\usepackage{algorithmicx}%
\usepackage{algpseudocode}%
\usepackage{listings}%
\usepackage{xcolor}
\usepackage{soul}
\usepackage{makecell}
\usepackage{gensymb}
\usepackage{mathtools}
\usepackage{setspace}
\usepackage{authblk}
\usepackage{wasysym}
\usepackage{physics}
\usepackage{url}
\usepackage{caption}
\usepackage{float}
\usepackage[section]{placeins}
\usepackage{cite}

\newcommand{\bl}[1]{\boldsymbol{#1}}

\title{Mixing Performance of Toroidal Ring Mixers: Effects of Flow Rate Ratios and Geometric Asymmetry}
\author[1]{Mohammad Majidi}
\author[1]{Taehong Kim}
\author[1]{Jiaqi Li}
\author[1]{Dongjie Jia}
\author[1]{Ehsan Rahimi}
\author[1,2,3]{Pavlos P. Vlachos}
\author[1]{Arezoo M. Ardekani\thanks{Correspondence: ardekani@purdue.edu}}

\affil[1]{School of Mechanical Engineering, Purdue University, West Lafayette, Indiana, USA}
\affil[2]{Weldon School of Biomedical Engineering, Purdue University, West Lafayette, Indiana, USA}
\affil[3]{Eli Lilly and Company, Indianapolis, Indiana, USA}
\begin{document}

\maketitle

\begin{abstract}

Microfluidic mixing is important for nanoparticle fabrication, where rapid contact between the solvent and nonsolvent streams is needed to control the formation process. Various micromixer geometries have been developed and analyzed to improve mixing efficiency.  However, for toroidal micromixers, the role of flow rate ratio and geometric asymmetry has not been examined in detail. In this study, the mixing process of two toroidal micromixer designs is investigated,   namely symmetric and asymmetric, with emphasis on the impact of flow rate ratio and geometric asymmetry during the mixing of miscible fluids. Numerical simulations are carried out to examine the mixing behavior of these toroidal micromixers for different flow rates and flow rate ratios. High-fidelity numerical simulations are performed using the stabilized finite element method. The concentration and velocity fields are used to  examine how the chamber asymmetry can affect the mixing performance. Experiments are also conducted to provide a validation for the numerical results. We demonstrate that the asymmetric toroidal mixer design generally improves the mixing over the conventional design, especially at low to moderate total flow rates.   The results obtained show that improved mixing can be achieved without changing the overall mixer size. 

\end{abstract}

\section{Introduction}

Microfluidic systems, which manipulate small volumes of fluids in channels with dimensions ranging from tens to hundreds of micrometers, have gained much attention recently in drug delivery, biochemical analysis, and nanoparticle synthesis \cite{nunes2022introduction}. For these systems, micromixers are important devices designed to promote mixing at small length scales \cite{stone2004engineering, beebe2002physics}. One of the main challenges with these devices is  their performance at low Reynolds numbers, where molecular diffusion is the dominant mixing mechanism. Since the diffusion  rate is slow over the practical channel lengths, achieving rapid and uniform mixing becomes difficult  \cite{stone2004engineering}. Various mixing techniques have been proposed and categorized into active and passive micromixers \cite{BAYAREH2020107771, hessel2005micromixers}. In active micromixers, mixing is assisted by an external input, such as an electric, magnetic, acoustic, or pressure field. In passive micromixers, the channel geometry itself is designed to stretch and rearrange the fluid streams to improve mixing \cite{hossain2009evaluation,  huang2013acoustofluidic, chen2023micromixing, dehghani2020mixing}.

Several passive micromixers have been proposed to enhance mixing in microfluidic systems. Some examples include split  and recombine micromixers \cite{afzal2012passive, raza2018effective}, staggered herringbone micromixers \cite{ansari2007shape, williams2008practical}, and Dean vortex-based or toroidal micromixers \cite{howell2004design, ligrani1988flow}. Their main advantage is improved mixing without external actuation or moving parts, which makes them simple, compact, and easier to integrate with other microfluidic components.  Among passive designs, toroidal and ring-shaped micromixers have received attention because they can enhance mixing through curved flow paths and repeated fluid rearrangement. \cite{mouza2008mixing,sheu2012mixing,alam2013mixing}.

 For these types of micromixers investigated in this study, many existing studies have relied on dye visualization, velocity measurement, or combined experimental and numerical approaches to understand the mixing process \cite{mouza2008mixing,ansari2010mixing,sheu2012mixing,alam2013mixing,ripoll2022optimal}. For example, Afzal and Kim numerically studied a passive micromixer with convergent--divergent sinusoidal channel walls and showed that stronger secondary flows improve mixing at moderate Reynolds numbers \cite{afzal2015optimization}. Mouza  et al.  experimentally studied a micromixer combining curved channels and split-and-recombine features and showed that these geometrical features reduce mixing time by generating three-dimensional flow patterns, especially at low Dean numbers \cite{mouza2008mixing}. Ansari and Kim investigated unbalanced split-and-recombine micromixers and showed that unbalanced geometries can improve mixing compared with balanced collisions \cite{ansari2010mixing}. Sheu  et al.  studied curved microchannels with split-and-recombine flow paths and showed that Dean vortex flow and uneven splitting increase the interface area between the two fluids and improve mixing \cite{sheu2012mixing}. Alam and Kim numerically studied a planar micromixer with circular chambers and split-and-recombine constriction channels. They showed that, at low Reynolds numbers, the split-and-recombine process increases the diffusion contact area and improves mixing, while at higher Reynolds numbers symmetric recirculating flows can reduce the mixing performance \cite{alam2013mixing}. More recently, Ripoll  et al.  experimentally investigated a toroidal micromixer for LNP fabrication over a wide range of flow rates using dye visualization and LNP size measurements \cite{ripoll2022optimal}. Their results showed that the imposed flow rate affects both the mixing process and the final LNP size, which reveals the importance of understanding the flow dynamics inside toroidal micromixers.

Although these studies have provided useful information on mixing performance in ring-shaped SAR and toroidal micromixers, the effect of flow-rate ratio and geometric asymmetry in toroidal micromixers has not been fully examined at the flow rates commonly used in experimental nanoparticle fabrication. In experiments, the internal three-dimensional flow and concentration fields inside the toroidal chambers are difficult to capture completely, especially when time-dependent or disturbed flow structures appear at higher flow rates. Computational fluid dynamics (CFD) is useful in this context because it provides detailed information on the flow and concentration fields inside the mixer, which are difficult to capture fully in experiments.

In this  numerical study validated by experiments, we investigate the mixing performance of two toroidal micromixer designs: a conventional design with symmetric toroidal chambers and a modified design with asymmetric toroidal chambers. In the symmetric design, the toroidal chambers are made of concentric annual channels, leading to a nearly uniform flow cross-section. In the asymmetric design, the chambers are eccentric annulus, resulting in a nonuniform flow cross-section that changes the flow path and alters the interaction between the two streams. We show that this asymmetric design will improve mixing while keeping the same overall mixer size. Streamlines, velocity vector fields, and mixing index are used to compare the mixing performance of the two designs over a range of flow rates and at two flow rate ratios.

This article is organized as follows.  we will first summarize the governing equations , numerical implementation, and simulation setup. Then, the numerical results and mixing performance of the two designs will be presented and compared with the experimental validation. Finally, we will summarize the findings and conclude the study.

\section{Methods}
\label{sec:methods}

This section describes the numerical and experimental procedures used in this study. First, the study design and simulation setup are introduced. The governing equations, numerical methods, and mesh generation are then described. Next, the experimental procedure is presented. Finally, the main definitions used in the analysis, including the mixing index and Reynolds number, are introduced.

\subsection{Study design and setup}

\subsubsection{Micromixer designs}
 
\begin{figure*}[!htbp]
    \centering
    \includegraphics[width=0.98\textwidth, trim={0cm 0cm 0cm 0cm},clip]{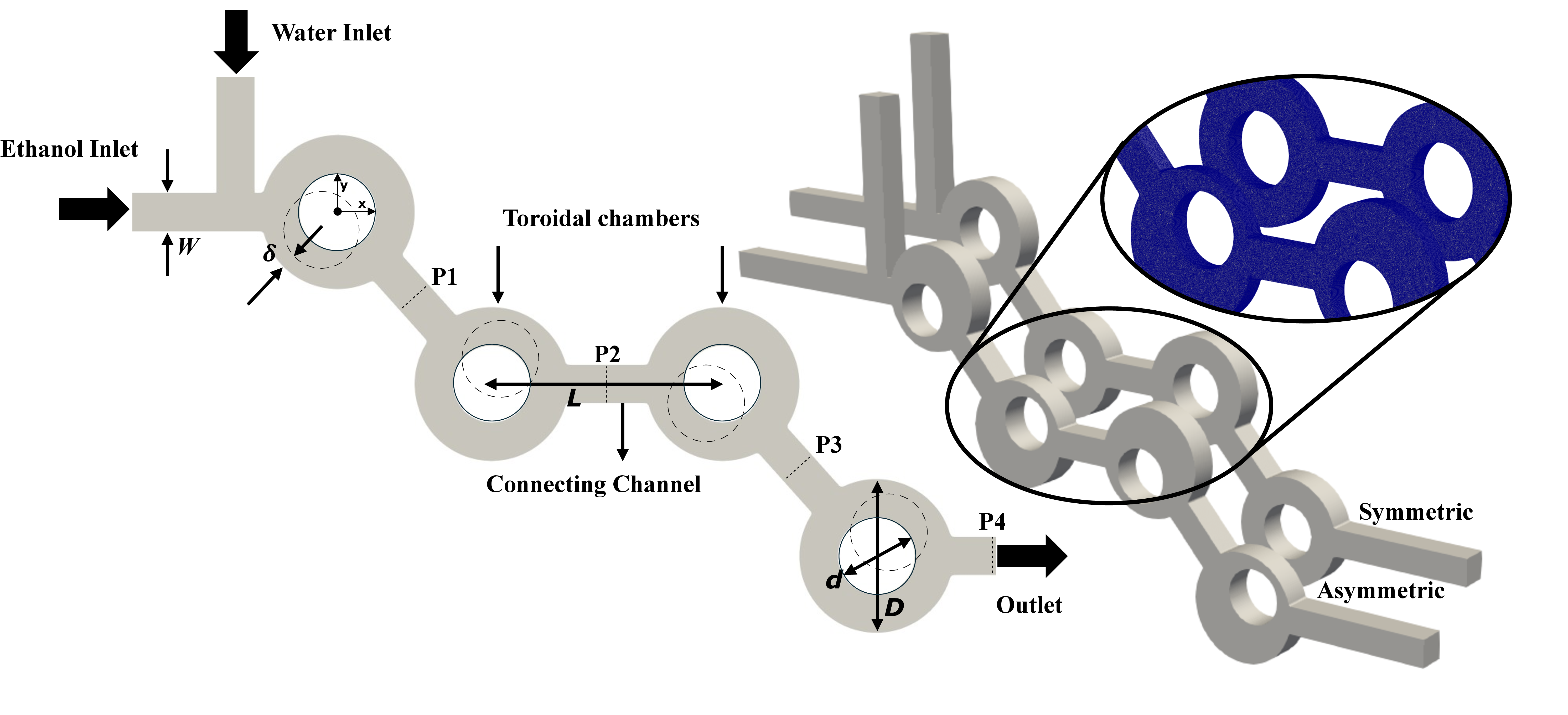}
    \caption{Schematic representation of the toroidal micromixer with labeled components, including the ethanol and water inlets, toroidal chambers, connecting channels, and outlet. Points P1--P3 represent the centers of the connecting channels, and P4 represents the outlet region used for mixing performance evaluation. The inset shows a representative view of the computational mesh.}
    \label{fig:setupgeometry}
\end{figure*}

Fig.~\ref{fig:setupgeometry} shows the two toroidal micromixer designs. In the symmetric design, the inner circle of each chamber is concentric with the outer circle. In the asymmetric design, the inner cores are shifted by about $130^\circ$ to the horizontal, which causes the flow cross-section to vary inside the flow channel. Aside from this difference, the two designs have the same channel height, width,  connecting channel sizes, and ring diameters. The full geometric details are listed in Table~\ref{tab:toroidal_mixer_design}.

\begin{table}[h]
    \centering
    \renewcommand{\arraystretch}{1.2} 
    \setlength{\tabcolsep}{10pt} 
    \caption{Geometrical parameters of the toroidal micromixer.}
    \begin{tabular}{|c|l|c|}
        \hline
        \textbf{Symbol} & \textbf{Description} & \textbf{Value ($c m$)} \\ \hline
        $W$  & Width of the channel & 0.015  \\  
        $D$  & Diameter of the toroidal chamber & 0.06 \\  
        $d$  & Diameter of the inner circular structure & 0.03 \\  
        $L$  & Pitch length (center-to-center distance between chambers) & 0.06 \\  
        $h$  & Height of the channel & 0.015 \\  
        $\delta$ & Gap size   &  
        \makecell{0.015(max) \\ 0.005 (min) }\\
        \hline
    \end{tabular}
    \label{tab:toroidal_mixer_design}
\end{table}

\subsubsection{Flow conditions}
Table~\ref{tab:flowrates} summarizes all cases considered in this study, which lists the ethanol flow rate $Q_e$, water flow rate $Q_w$, and total flow rate $Q$ for each case.  Two flow rate ratios (FRRs) are considered: $\mathrm{FRR}=2{:}2$, representing the balanced flow case, and $\mathrm{FRR}=1{:}3$, representing the unbalanced flow case.  These flow rates are chosen  based on common operating conditions in drug manufacturing setting. 

 \begin{table}[h!]
\centering
\caption{Flow conditions considered in this study}
\label{tab:flowrates}
\begin{tabular}{cccc}
\hline
FRR $(Q_e:Q_w)$ & $Q_e$ (mL/min) & $Q_w$ (mL/min) & $Q$ (mL/min) \\
\hline
2:2 & 0.5  & 0.5  & 1 \\
2:2 & 1.0  & 1.0  & 2 \\
2:2 & 2.0  & 2.0  & 4 \\
2:2 & 4.0  & 4.0  & 8 \\
\hline
1:3 & 0.25 & 0.75 & 1 \\
1:3 & 0.50 & 1.50 & 2 \\
1:3 & 1.00 & 3.00 & 4 \\
1:3 & 2.00 & 6.00 & 8 \\
\hline
\end{tabular}
\end{table}

\subsection{Computational fluid dynamics}
The incompressible Navier-Stokes equations are written on the computational domain, $\Omega$, for all $t \in (0,T)$ as
\begin{equation}
    \begin{alignedat}{3}
    \rho \frac{\partial \bl u}{\partial t} + \rho \bl u \cdot \nabla{\bl u} + \nabla p - \nabla\cdot\boldsymbol{\tau} = 0,  \\
    \nabla\cdot\bl u  = 0,  \\
    \end{alignedat}
 \label{eqn:NS-og}
\end{equation}
where $\bl u$ is the velocity vector, $p$ is the pressure, and $\rho$ is the mixture density. The deviatoric stress tensor, $\boldsymbol{\tau }$, is expressed using the Newtonian constitutive equation as
\begin{equation}
    \boldsymbol {\tau }=\mu \left[\nabla \mathbf{u} +(\nabla \mathbf {u} )^{\intercal }\right],
    \label{eqn:tau_stress}
\end{equation}
where $\mu$ is the mixture dynamic viscosity. 

The nonuniform and nonconstant mixture density, $\rho$, and dynamic viscosity, $\mu$, are calculated as
\begin{equation}
    \begin{alignedat}{3}
    \rho &= c\rho_e + (1-c)\rho_w, \\
    \mu  &= \exp \left[c\ln{\mu_e} + (1-c)\ln{\mu_w} + c(1-c)d_{ew}\right], \\
    \end{alignedat}
 \label{eqn:rho_mu}
\end{equation}
where the subscripts $e$ and $w$ denote ethanol and water, respectively, and $c$ tracks the concentration of ethonal in the fluid mixture.
The Grunberg-Nissan model is for the mixture viscosity, where the empirical binary interaction coefficients, $d_{ew}=3.6$, are approximated from experimental data~\cite{furukawa2017brownian}.

The evolution of the concentration, $c$, is governed by the unsteady advection-diffusion equation, written on the computational domain, $\Omega$, for all $t \in (0,T)$ as
\begin{equation}
    \frac{\partial c}{\partial t} + \bl u \cdot \nabla{c}  = D\nabla^2c,
\label{eqn:AD-og}
\end{equation}
where $c$ is the volume fraction of ethanol, and $D$ is the diffusivity of ethanol in water. 

The above governing equations are implemented in an in-house finite element solver named Multiphysics Finite Element Solver (MUPFES)~\cite{moghadam2013modular,esmaily2015bi,esmaily2013new}. 
The solver has been extensively validated and deployed to simulate turbulent flows in various studies~\cite{jia2022characterization,jia2021simulation,jia2023time,jia2026turbulent}. 
The exact implementation and solution procedure used in this study can be found in our previous publication~\cite{jia2025high}.

In summary, The Navier-Stokes equations are solved using the residual-based variational multiscale method (RBVMS)~\cite{bazilevs2007variational,bazilevs2008isogeometric}. The advection-diffusion equation is solved using the Streamline upwind Petrov–Galerkin (SUPG) method~\cite{brooks1982streamline}. 
The solver uses a second-order generalized-$\alpha$ method for the time integration~\cite{jansen2000generalized}, where $\rho_{\infty} = 0.2$.
Linear shape functions are used for all mesh elements. 
The nonlinear systems of equations are linearized and iteratively solved using a modified Newton-Raphson method.
The linear system for the Navier-Stokes equations is solved using a bi-partitioned iterative algorithm specifically designed for solving the linear system from the Navier-Stokes equations~\cite{esmaily2015bi}.
The linear system for the advection-diffusion equation is solved using a preconditioned generalized minimal residual (GMRES) algorithm~\cite{shakib1989multi}.

Throughout this study, ethanol and water are used as the two mixing liquids in all simulations. Their room-temperature properties are taken as $\rho_e = 0.789\;\mathrm{g/cm^3}$ and $\mu_e = 1.2\;\mathrm{cP}$ for ethanol, and $\rho_w = 1\;\mathrm{g/cm^3}$ and $\mu_w = 1\;\mathrm{cP}$ for water. The diffusivity of ethanol in water is to $D = 1.23\times10^{-5}\;\mathrm{cm^2/s}$. 

 Unstructured tetrahedral meshes were generated using TetGen \cite{hang2015tetgen} for both the symmetric and asymmetric designs. A  maximum element  volume  constraint was specified to  produce  around 6 million tetrahedral elements for both designs. This level of mesh was chosen to provide a sufficient description of the flow field inside both mixers. The final meshes contain $6{,}063{,}390$ tetrahedral elements for the symmetric design and $6{,}063{,}407$ tetrahedral elements for the asymmetric design.

\subsection{Experimental procedure}
 
\subsubsection{Fabrication of the device}

 After defining the numerical cases, the corresponding microfluidic devices were fabricated for experimental comparison.  The microfluidic master mold was fabricated via standard photolithography. In detail, a silicon wafer was spin-coated with AZ-10XT 520cp photoresist (Merck), exposed using a maskless aligner (MLA 150, Heidelberg), and developed. The pattern was etched by deep reactive ion etching (Versaline DSE, Plasma-Therm), followed by silanization using trichloro(1H,1H,2H,2H-perfluorooctyl)silane (97\%, Sigma-Aldrich) via chemical vapor deposition (CVD). PDMS (Sylgard 184, Dow) was mixed at a 10:1 base-to-curing-agent ratio, degassed, poured onto the mold, and cured overnight at 70\textdegree C. Inlet and outlet holes were punched, and plastic connectors were inserted as chip-to-world fluidic interfaces. For assembly, the PDMS channel layer was plasma-bonded to a flat PDMS substrate. After plasma treatment at the top of the channel layer, uncured PDMS was poured over the device and cured to form a top layer in place. The final device consists of three plasma-bonded PDMS layers: a substrate, a micromixer layer, and an integrated capping layer for structural reinforcement, which provides enhanced mechanical strength and bonding integrity, enabling reliable operation under high flow rates.

\subsubsection{Micro-LIF measurements}

To complement the simulation, we conducted micro-laser induced fluorescence ($\upmu$-LIF) experiments under the same total flow rates and flow-rate ratios considered in the numerical study. The experiments were carried out on a Nikon Eclipse Ti microscope at 4$\times$ magnification. Rhodamine B (excitation peak at 545 nm and emission peak at 566 nm) was added to ethanol to achieve a concentration of 100 $\upmu\mathrm{M}$, while DI water was left untreated. A dual-channel syringe pump was used for the two inlet streams. Fluorescence images were acquired using a Nikon Eclipse Ti microscope with a $4\times$ objective, Andor Zyla CMOS camera, and 89 North Laser Diode Illuminator (LDI) for illumination. The exposure time was 100 ms, and the illumination intensity (10\%) and camera gain were kept constant for all cases. For each flow condition, 30 images were acquired after the flow reached a statistically steady state.

\subsubsection{Micro-PIV measurements and processing}

Micro-particle image velocimetry ($\upmu$PIV) was used to measure the in-plane velocity field at the channel center plane. The flow was seeded with fluorescent polystyrene tracer particles with a diameter of 2 $\upmu$m at a concentration of 0.005\%. The particles were illuminated using the same laser used for LIF experiment, and image pairs were acquired using a specialized PIV camera (IMPERX) mounted on the Nikon Eclipse Ti microscope with a 10$\times$ objective. The measurement plane was positioned at $z/h=0.5$, corresponding to the channel center plane. The pixel resolution was 0.55 $\upmu\mathrm{m/pixel}$.

For each condition, 500 image pairs were acquired after the flow reached a statistically steady state. The inter-frame time $\Delta t$ was set to 10 $\upmu\mathrm{s}$ based on the flow rate to maintain particle displacements within the optimal range for cross-correlation. The image pairs were processed using a ensemble-averaging multi-pass cross-correlation algorithm with a final interrogation windows of 32 $\times$ 32 pixels, with 75\% overlap. Spurious vectors were removed using median filter validation and replaced by interpolation from neighboring valid vectors.

\subsection{Mixing characteristics}

To quantify the degree of mixing, the mixing index (MI) is defined as
\begin{equation}
\mathrm{MI} = 1 - \sqrt{\frac{\sigma^{2}}{\sigma_{\max}^{2}}},
\end{equation}
where $\sigma^{2}$ represents the variance of the concentration field and $\sigma_{\max}^{2}$ denotes the maximum possible variance for the given inlet concentration ratio. The value of $\sigma_{\max}^{2}$ depends on the FRR between the two inlet streams. For the cases considered here, $\sigma_{\max}^{2}=0.25$ for $\mathrm{FRR}=2:2$, and $\sigma_{\max}^{2}=0.1875$ for $\mathrm{FRR}=1:3$. 
The MI value ranges from 0, corresponding to immiscible fluids, to 1, which indicates a perfectly homogeneous mixture. The mixing index was calculated once the flow field and concentration field achieved a steady state. Hence, the reported values represent the final mixing behavior rather than transient effects. Total flow rates $Q = 1~\mathrm{mL/min}$ and $Q = 2~\mathrm{mL/min}$ result in time-steady laminar flows, whereas   $Q = 4~\mathrm{mL/min}$ and $Q = 8~\mathrm{mL/min}$ result in unsteady transitional flow. For the unsteady flows, the values of the mixing index were averaged over time.

In this study, the variance $\sigma^{2}$ is computed from the concentration field $c$ over a cross-sectional area $S$ as
\begin{equation}
\sigma^{2} =
\frac{\int_{S} (c-\bar{c})^{2}\, dS}{\int_{S} dS},
\end{equation}
where $\bar{c}$ is the spatially averaged concentration over the same cross-section, defined as
\[
\bar{c} = \frac{\int_{S} c\, dS}{\int_{S} dS}.
\]

For the experiments, the mixing index was calculated using the fluorescence intensity profile instead of the numerical concentration field. Since the experimental images are only two-dimensional, the intensity profile of the image, $I(x)$, was taken along a line crossing the channel at each measurement location. The experimental mixing index, $MI_{\text{exp}}$, is then defined as 

\begin{equation}
MI_{\text{exp}}=1-\sqrt{\frac{\frac{1}{W}\int_0^{W}\left(I(x)-\overline{I}\right)^2\, dx}{\sigma_{\max}^2}},
\end{equation}
where $W$ is the width of the channel, $\overline{I(x)}$ is the mean intensity along the selected line. For each experimental condition, $MI_{\text{exp}}$ was first calculated for each individual image and then averaged over all images in a run.

 In this study, the Reynolds number is defined as follows:
\begin{equation}
\mathrm{Re} = \frac{U\,h}{\nu_e},
\end{equation}
where $U$ is the characteristic velocity defined as the total flow rate divided by the outlet channel cross-sectional area, $h$ is the channel height, and $\nu_e$ is the kinematic viscosity of ethanol    ($\nu_e$=$\mu_e$/$\rho_e$), which is used as the reference kinematic viscosity. For the flow conditions considered here, Reynolds number ranges from $\mathrm{Re} = 100$ to $\mathrm{Re} = 900$.

\section{Results } 
\label{sec:results}

 This section presents the mixing process in the symmetric and asymmetric toroidal micromixers. First, the mixing performance at the outlet is compared for the two designs under different total flow rates and FRRs. Then, the evolution of the concentration field along the mixer is discussed to show how the mixing process evolves. Streamlines and velocity vector fields are used to explain the difference between the two designs. Finally, the numerical results are compared with the experimental measurements to validate the results.

\begin{figure*}[!htbp]
    \centering
    \includegraphics[width=0.8\textwidth, trim={0cm 0cm 0cm 0cm},clip]{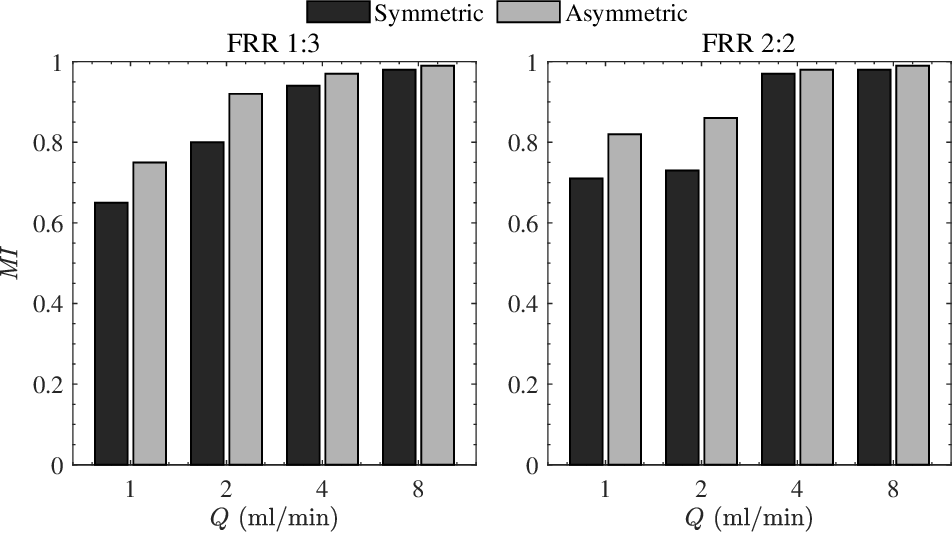}
    \caption{Comparison of the mixing index ($\mathit{MI}$) for symmetric and asymmetric mixer designs across different total flow rates ($Q$) for FRRs~1:3 and 2:2.
}
    \label{fig:mivsQ}
\end{figure*}

The mixing performance at the outlet for both mixer geometries is shown in Fig.~\ref{fig:mivsQ} based on all of the chosen total flow rates and FRRs.  Overall, the results show that increasing the total flow rate  improves the mixing performance, although the level of this improvement depends on both the design and the FRR. For the balanced flow when $\mathrm{FRR}=2{:}2$, the asymmetric design consistently performs better  than the symmetric design for all the total flow rates, with the most noticeable improvement observed at $Q = 1$ and $2~\mathrm{mL/min}$.  A similar trend is also observed for $\mathrm{FRR}=1{:}3$, although the improvement in mixing is less pronounced than that seen for $\mathrm{FRR}=2{:}2$  cases. Overall, these observations are in good agreement with the experimental results reported by Ripoll et al.~\cite{ripoll2022optimal}  for the symmetric design, who considered similar total flow rates for $\mathrm{FRR}=1{:}3$ and obtained mixing index values close to those found in Fig.~\ref{fig:mivsQ}.

\begin{figure*}[!htbp]
    \centering
    \includegraphics[width=0.7\textwidth, trim={0cm 0cm 0cm 0cm},clip]{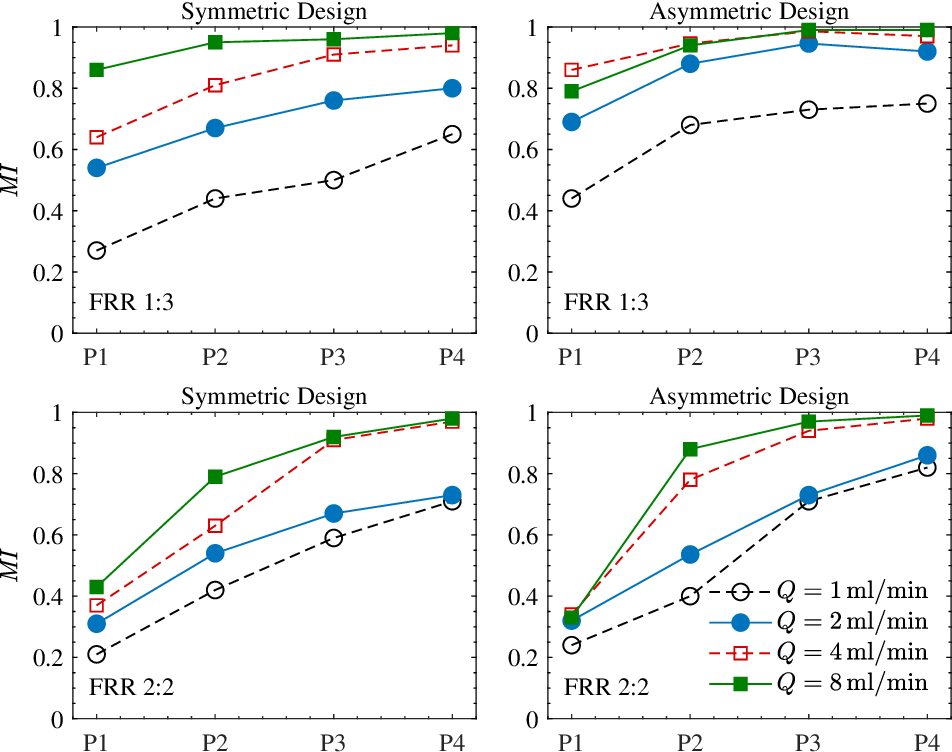}
    \caption{Comparison of the mixing index ($\mathit{MI}$) evolution at different positions for symmetric (left) and asymmetric (right) mixer designs under various total flow rates and flow-rate ratios (FRR). Each row corresponds to a different FRR. The mixing index is evaluated at four locations along the mixer, where $P1$--$P3$ denote the centers of successive constriction channels, and $P4$ represents the outlet region.
}
    \label{fig:mivsx}
\end{figure*}

Fig.~\ref{fig:mivsx}  evaluates the mixing along the  mixers to further highlight the effect of geometric asymmetry on mixing performance at four locations. Points P1 to P3 are located at the center of the connecting channels, while P4 is located at the outlet area, as shown in Fig. \ref{fig:setupgeometry}.  Overall, the asymmetric design leads to both earlier mixing homogenization ($MI > 0.8$) in the device and higher degree of mixing at the outlet. The  exact degree of improvement depends on the FRR and the imposed total flow rate. In particular, this improvement in mixing process is  most obvious when $\mathrm{FRR}=1{:}3$. As the flow advances downstream, the  mixing index value at P3, where the mixed flow leaves the third chamber, is almost equal to its value at the outlet. As the total flow rate increases,  the mixing index converges to unity after the second chamber.  For the unbalanced flow case, the asymmetric design can achieve nearly the same outlet mixing quality with fewer chambers, reducing the required mixing length and residence time, which is desirable for LNP fabrication \cite{evers2018state}.

\begin{figure*}[!htbp]
    \centering
    \includegraphics[width=0.8\textwidth, trim={0cm 0cm 0cm 0cm},clip]{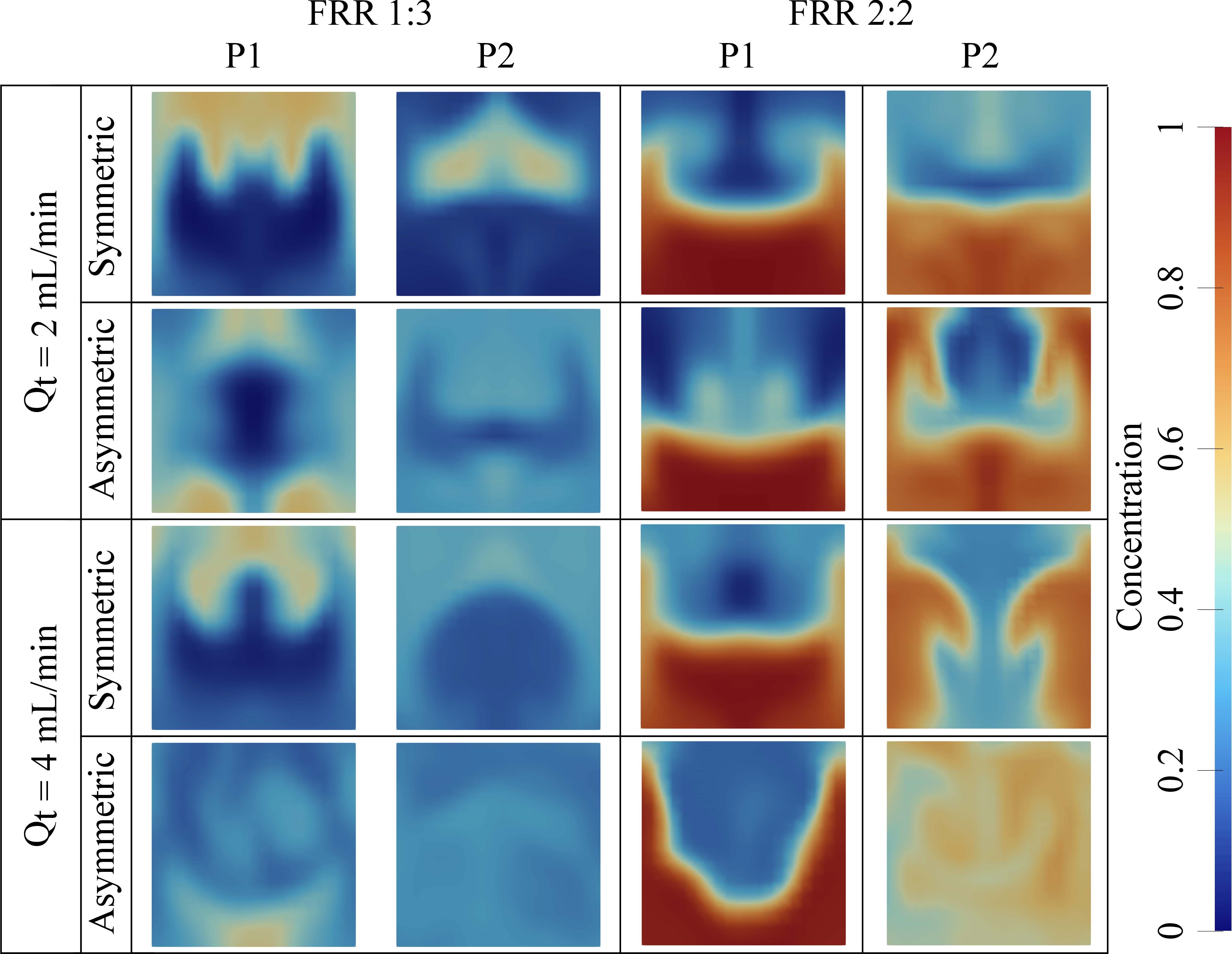}
    \caption{Concentration field distributions at the first two cross-sectional locations, P1 and P2, at $Q_t=2 \text{ and } 4\text{ mL/min}$ for both FRRs. The first two columns are for the unbalanced flow ($\mathrm{FRR}=1{:}3$), and the second two columns are for the balanced flow ($\mathrm{FRR}=2{:}2$). }
    \label{fig:mixingvsxfr13}
\end{figure*}

 To better understand the faster  mixing in the asymmetric design, the concentration fields are shown at $Q_t=2$ and $4~\mathrm{mL/min}$ for both FRRs in Fig.~\ref{fig:mixingvsxfr13}.  The contours are taken at P1 and P2 to explain the early difference of mixing mechanism between the two designs .   For laminar micromixers, mixing is improved by increasing the contact area between the two fluid streams \cite{hessel2005micromixers}. 
 
  For the unbalanced flow case, $\mathrm{FRR}=1{:}3$, the symmetric design shows a more separated concentration pattern at P1 and P2, while the asymmetric design shows a wider contact region between the two streams at these locations. Since the $1{:}3$ flow already has flow asymmetry because of the unequal inlet flow rates, the improvement from the geometry asymmetry is less significant. However, for the balanced flow case, $\mathrm{FRR}=2{:}2$, the effect of geometry asymmetry is more clear because the two inlet streams enter with equal flow rates. In the symmetric design, the two streams remain relatively separated at P1 and P2, and the interface between them remains relatively smooth and less stretched. In contrast, the asymmetric design introduces an imbalance in the flow and causes a larger contact region to get generated between the two fluids at these locations. This, in turn, drastically improves mixing as the mixed fluid flows toward the outlet area.   It is worth mentioning that for both FRRs, the asymmetric design at $Q_t=4~\mathrm{mL/min}$ also shows onset of  unsteady transitional flow behavior, which improves the mixing performance  through introduction of instability .

\begin{figure*}[!htbp]
    \centering
    \includegraphics[width=0.6\textwidth, trim={0cm 0cm 0cm 0cm},clip]{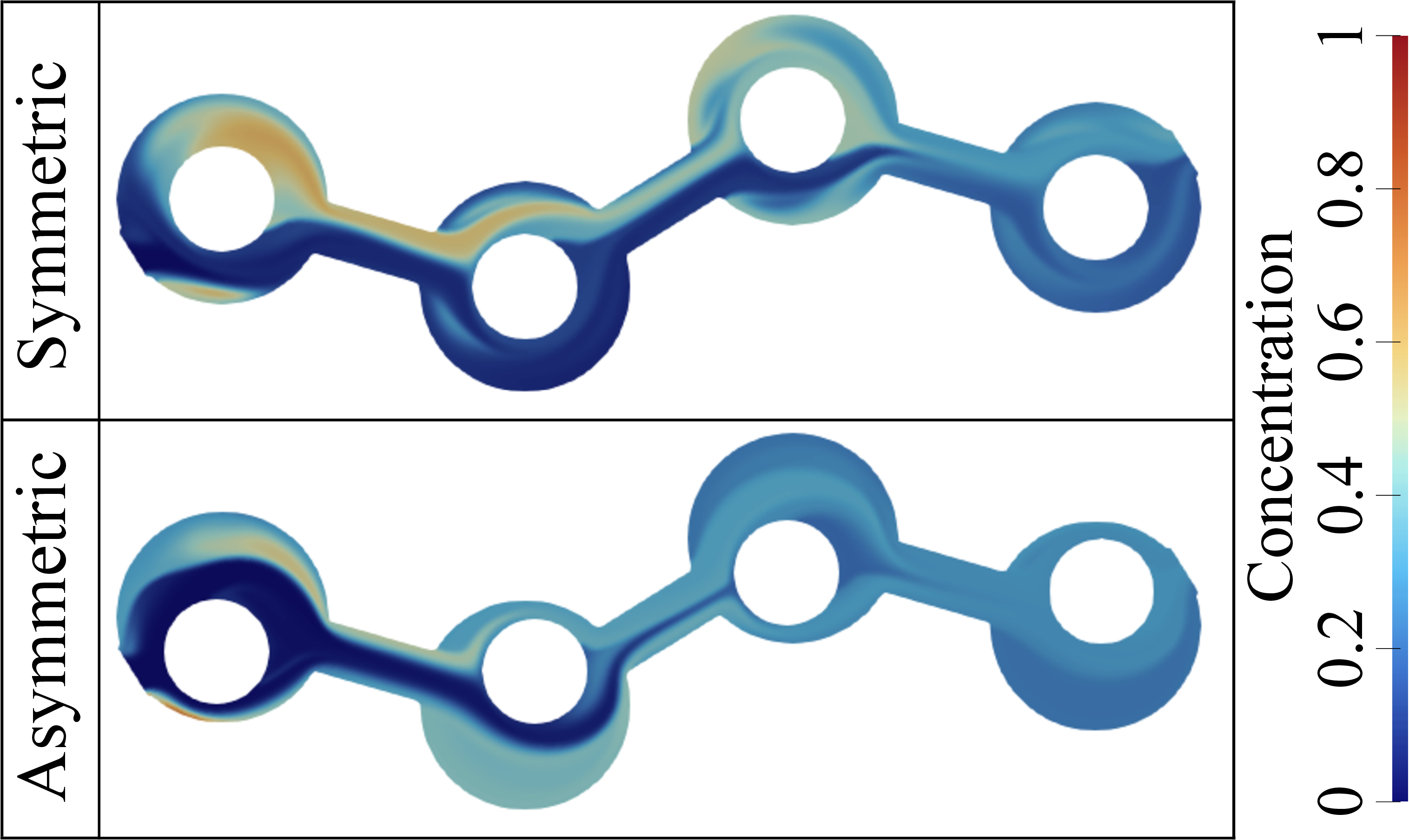}
    \caption{The comparison of concentration field distributions in the mid-plane of each chamber between symmetric (top row) and asymmetric (bottom row) designs at $Q_t=2 \text{ mL/min}$ for the unbalanced flow, $\mathrm{FRR}=1{:}3$ }
    \label{fig:c_13}
\end{figure*}

To show an example of how the two designs differ inside the chambers, Fig.~\ref{fig:c_13} shows the concentration fields along the mixer midplane for $\mathrm{FRR}=1{:}3$ at $Q_t=2~\mathrm{mL/min}$.  The symmetric design performs as a split-and-recombine mixer, resulting in little mixing inside the toroidal chambers themselves. In contrast,  the asymmetric design  forces the contact of the two fluid streams on the same side.   As a result , mixing is  complete  by the third chamber. In the symmetric design, however, the concentration field is still not uniform even in the fourth chamber.  This supports the previous observation that, for the unbalanced-flow case, the asymmetric design can reach near-outlet mixing quality with fewer chambers.

\begin{figure*}[!htbp]
    \centering
    \includegraphics[width=0.6\textwidth, trim={0cm 0cm 0cm 0cm},clip]{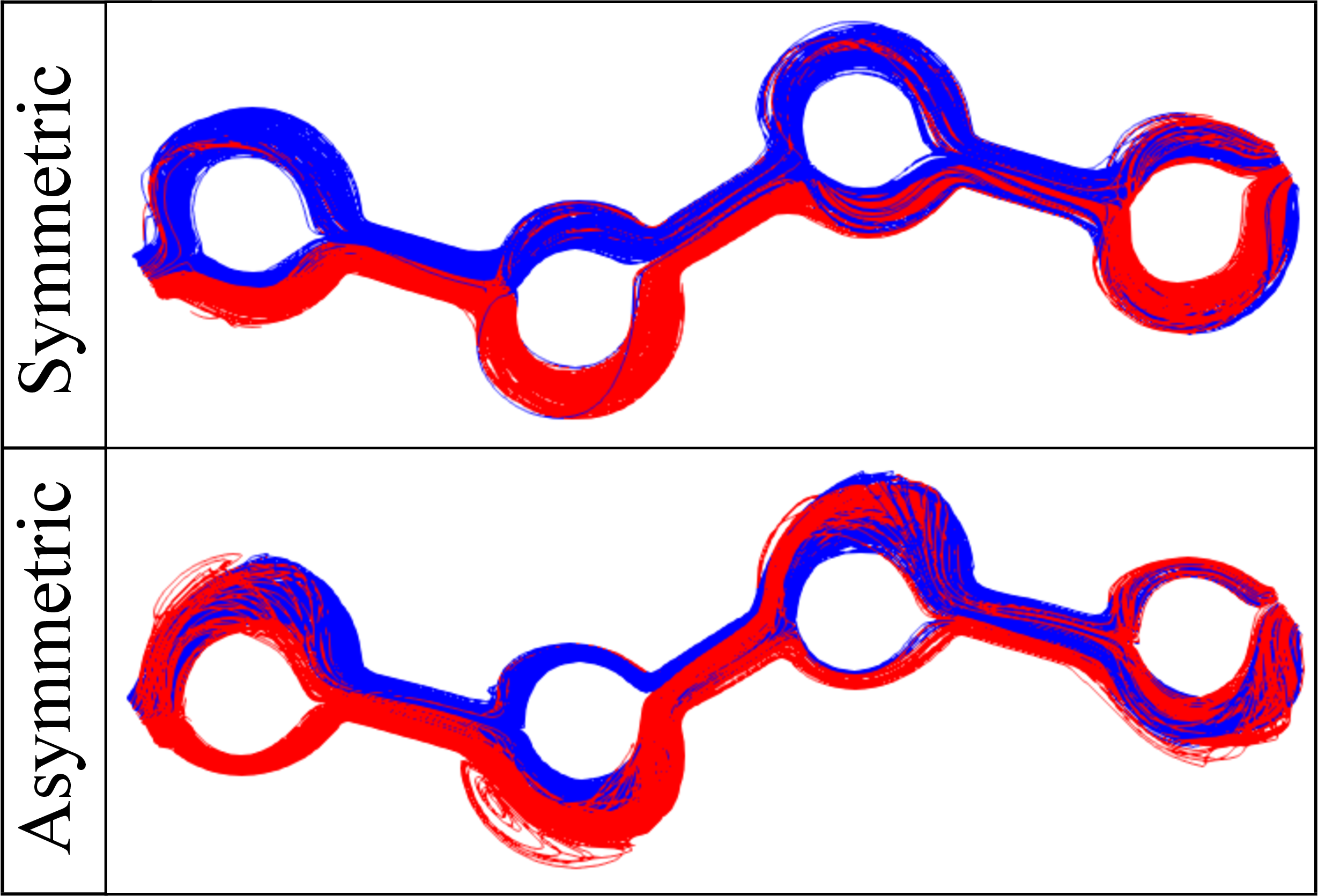}
    \caption{Comparison of the streamlines in the mixer chambers for symmetric (top row) and asymmetric (bottom row) designs at $Q_t = 2\,\mathrm{mL/min}$ for the balanced flow, $\mathrm{FRR}=2{:}2$. Blue streamlines correspond to water and red streamlines correspond to ethanol.}
    \label{fig:S_22}
\end{figure*}

 The streamline patterns in Fig.~\ref{fig:S_22} are shown at $Q_t=2~\mathrm{mL/min}$ for the balanced flow case, $\mathrm{FRR}=2{:}2$, where the difference between the two designs is clear. The red streamlines represent ethanol and the blue streamlines represent water. In the symmetric design, the two streams remain separated as they pass through the chambers, and the flow mainly behaves like a split-and-recombine mixer. This results in limited mixing inside the toroidal chambers themselves. In the asymmetric design, the two sets of streamlines are forced to contact on the same side of the chambers.

\begin{figure*}[!htbp]
    \centering
    \includegraphics[width=0.65\textwidth, trim={0cm 0cm 0cm 0cm},clip]{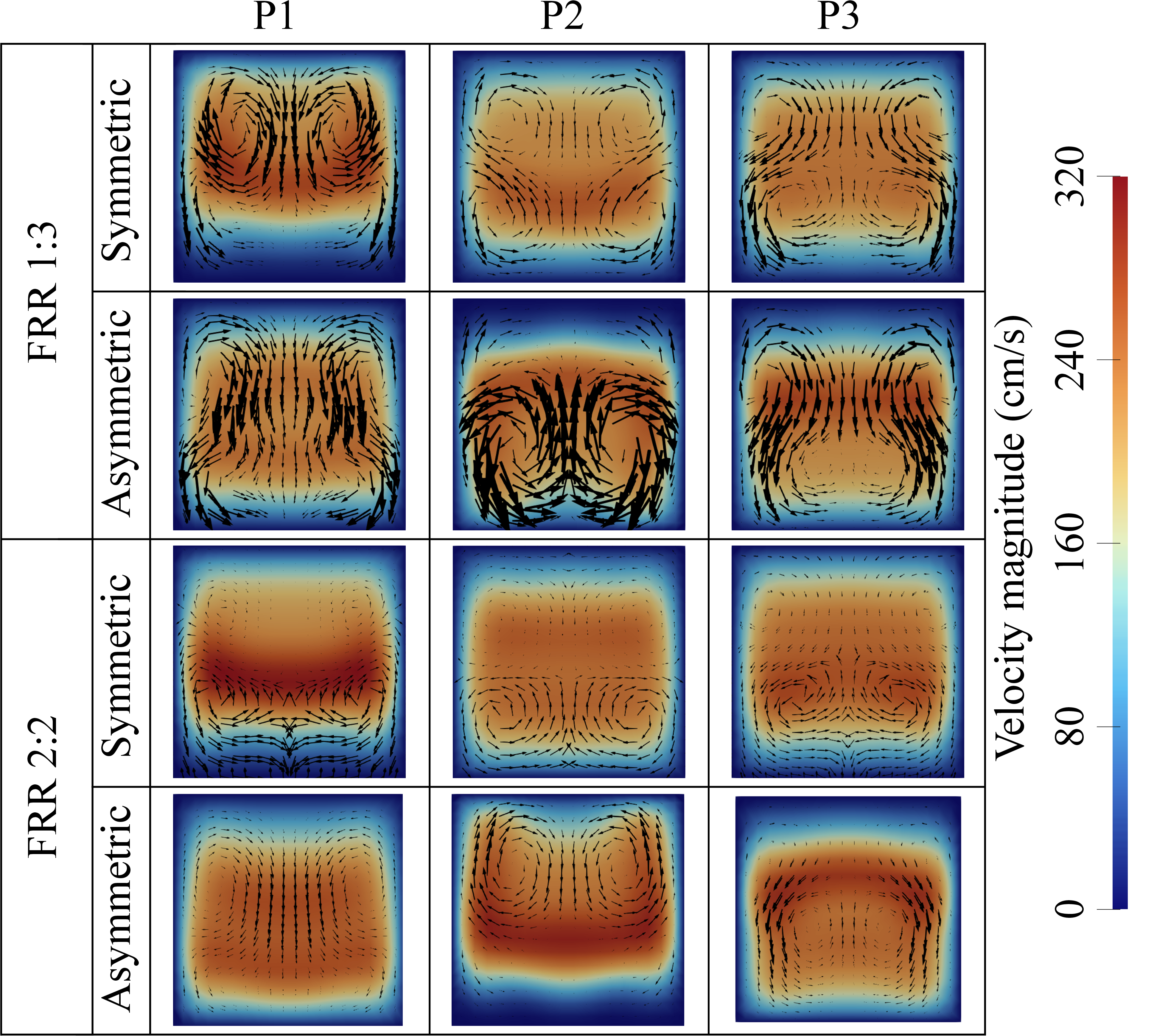}
    \caption{In-plane velocity vector fields at the centers of the constriction channels (P1–P3) for the symmetric and asymmetric mixer designs for the unbalanced ($\mathrm{FRR}=1{:}3$) and balanced ($\mathrm{FRR}=2{:}2$) flows at $Q = 2$ mL/min. .
}
    \label{fig:VV_13}
\end{figure*}

Fig.~\ref{fig:VV_13} demonstrate the velocity vectors at the cross sections located at the middle of the constriction channels from P1 to P3 at $Q_t=2~\mathrm{mL/min}$ for both FRRs.  These velocity vector plots are used to compare the strength of the spiraling motion in the connecting channels. For the unbalanced flow, $\mathrm{FRR}=1{:}3$, the difference between the two designs is most obvious. The asymmetric design shows stronger spiraling patterns and higher peak velocity vectors in these connecting channel cross sectional areas, which helps improve the fluids mixing. For the balanced flow, $\mathrm{FRR}=2{:}2$, the velocity vector patterns are fairly similar in the symmetric and asymmetric designs, although the asymmetric case still appears slightly stronger and shows slightly larger spiraling fluid motion.

\subsection{Experimental validation}
\label{sec:exp validation}

\begin{figure*}[!htbp]
    \centering
    \includegraphics[width=0.7\textwidth, trim={0cm 0cm 0cm 0cm},clip]{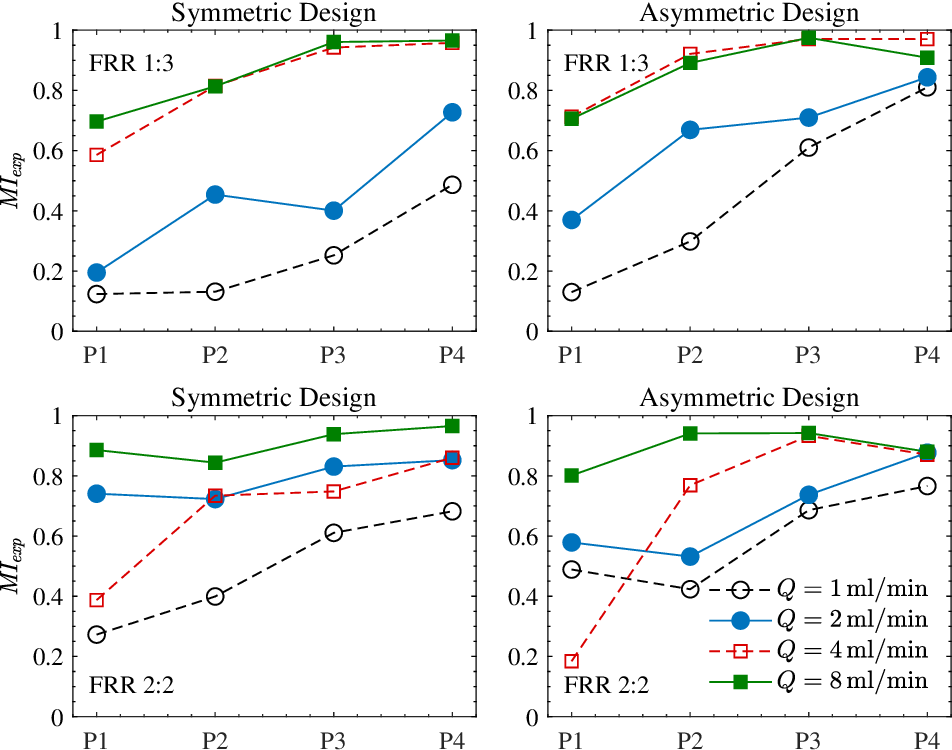}
    \caption{Experimental mixing index ($\mathit{MI}$) measured at different regions of the symmetric (left) and asymmetric (right) mixer designs under various total flow rates and flow-rate ratios (FRR). Each row corresponds to a different FRR. The experimental MI is evaluated over four regions along the device (P1 to P4), which correspond to the same measurement locations used in the simulation results shown in Fig. \ref{fig:mivsx}.
}
    \label{fig:mivsx_exp}
\end{figure*}

To compare the evolution of mixing along the channels, similar to Fig.~\ref{fig:mivsx}, we calculated the Mixing Index ,$MI_{\text{exp}}$, based on the intensity of the image \cite{ripoll2022optimal}. Four locations (P1, P2, P3, and P4) within the straight channel portion are selected after each ring, as informed by the simulation. The corresponding experimental results are shown in Fig.~\ref{fig:mivsx_exp}.  

Overall, the spatial progression of mixing index agrees with the numerical results. We observe an increasing $MI_{\text{exp}}$ along the channel, which also improves with the imposed total flow rate.  The evolution of $MI_{\text{exp}}$ not only highlights the improved mixing performance at moderate flow rates achieved by a simple change of the channel design, but also shows strong overall agreement with the simulation results for FRR of 1:3. While there is some mismatch between the experiment and simulation exists, we do not expect a perfect match considering the different definitions for the mixing indices. It's worth mentioning also that in the simulation, mixing index is defined directly using mixing concentration over the full cross-section of the 3D channel, whereas the experimental measurement is based on image intensity along a line near the center 2D plane.

\begin{figure*}[!htbp]
    \centering
    \includegraphics[width=0.98\textwidth, trim={0cm 0cm 0cm 0cm},clip]{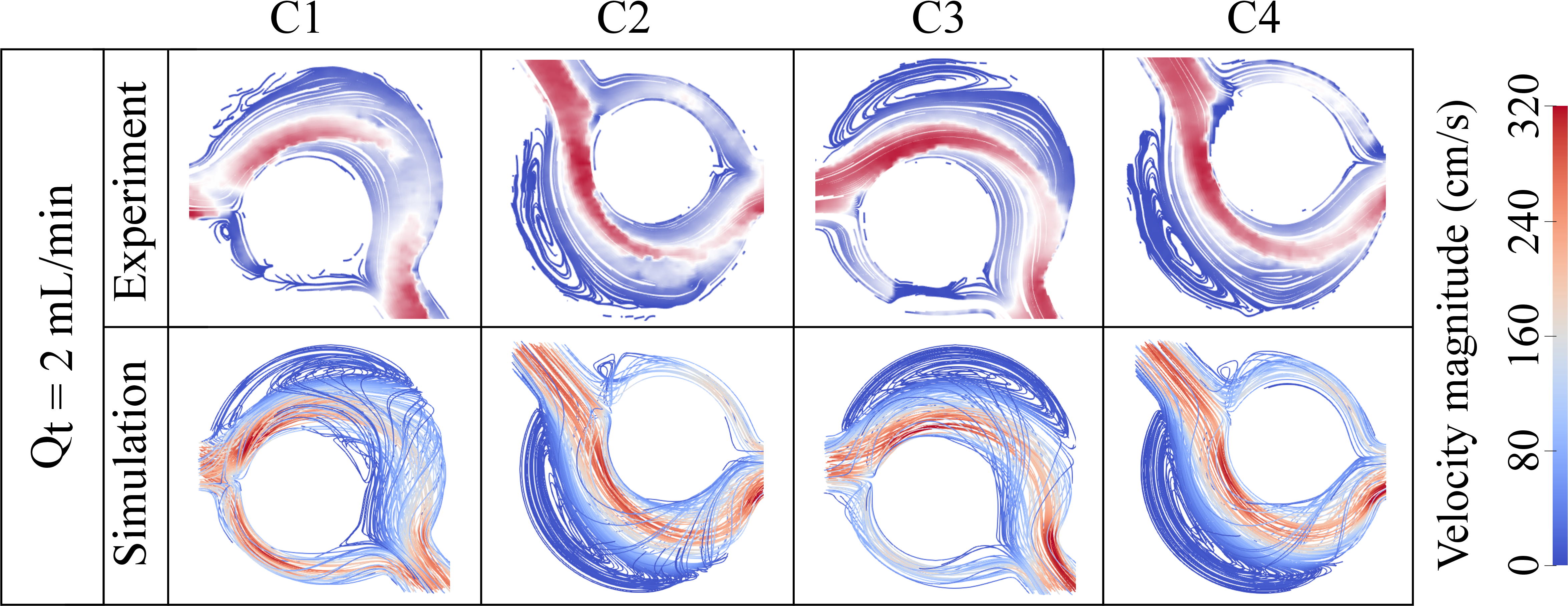}
\caption{Comparison of streamlines obtained from 2D PIV measurements and 3D numerical simulations for the asymmetric design at a flow rate of $Q = 2~\mathrm{mL/min}$ and $\mathrm{FRR} = 1{:}3$.}
    \label{fig:streamline_exp}
\end{figure*}

To further validate the results, we also compare the streamlines obtained from 2D PIV measurements with those from the 3D numerical simulations for the asymmetric design at a total flow rate of $Q = 4~\mathrm{mL/min}$ and $\mathrm{FRR}=1{:}3$, as shown in Fig.~\ref{fig:streamline_exp}. The overall flow topology, including the primary circulation patterns within the chambers, shows good qualitative agreement between the experimental and numerical results.

\section{Conclusion}
\label{sec:conclusion}

 In this study, the mixing performance of two toroidal micromixer designs, symmetric and asymmetric, was investigated using   high-fidelity CFD  simulations and complementary experiments.  The results showed that the asymmetric design generally provides better mixing performance than the symmetric design.   
 Along the mixer length, the asymmetric design  demonstrated  a faster  homogenization  than the symmetric design. This improvement is due to the shifted inner circles, which  breaks the flow symmetry  inside the chambers and linking channels.  This introduced asymmetry increases the contact area between the two fluids and triggers instability at a lower flow rate, both of which improve the mixing performance. The experimental results supported the numerical results. The measured MI along the channel length followed the same trend as the simulations. The chamber circulation patterns from the 2D PIV data were also similar to the 3D numerical results for the asymmetric design. In summary, using asymmetric chambers improved the mixing performance of the toroidal mixer without changing the overall mixer size or adding active components. Asymmetric mixer showed that it could reach near-outlet mixing quality before the final chamber, which suggests that the total  number of chambers  can be  reduced  for applications  with strict size constraints.

\section*{Data availability}
The simulation files and results presented in this study are available on request from the corresponding author.

\section*{Acknowledgment}

The authors would like to acknowledge that this study is supported by Eli Lilly and Company.  

\section*{Author contributions statement}
Conceptualization, M.M., T.K., J.L., A.M.A.; computational methodology, M.M., D.J., E.R. and A.M.A.; software, M.M. and D.J.; validation, M.M., J.L, D.J.; formal analysis, M.M., T.K, J.L.; investigation, M.M.; resources, A.M.A., P.V.; data curation, M.M., T.K., and J.L.; writing---original draft preparation, M.M., T.K, and J.L.; writing---review and editing, M.M., T.K., J.L., D.J., E.R., A.M.A., and P.V.; visualization, M.M., T.K., and J.L.; supervision, D.J., A.M.A., and P.V.; project administration, A.M.A. and P.V. All authors have read and agreed to the published version of the manuscript.

\bibliographystyle{unsrt}
\bibliography{ref}

\end{document}